# Manipulation of exciton and trion quasiparticles in monolayer WS$_2$ via charge transfer

Anand P. S. Gaur[1,2*], Adriana M. Rivera[1], Saroj P. Dash[3], Sandwip Dey[4], Ram S. Katiyar[1], Satyaprakash Sahoo[5,6*]

[1]Department of Physics and Institute for Functional Nanomaterials, University of Puerto Rico, San Juan, PR 00931 USA

[2]Department of Materials Science and Engineering, Iowa State University, Ames, IA 50011, USA

[3]Department of Microtechnology and Nanoscience, Chalmers University of Technology, SE-41296, Göteborg, Sweden

[4]Materials Science and Engineering in the School for Engineering of Matter, Transport, and Energy, Arizona State University, Tempe, Arizona 85287, USA

[5]Institute of Physics, Bhubaneswar 751005, India

[6]Homi Bhabha National Institute, Anushaktinagar, Mumbai 400 085, India

**Abstract:**

Charge doping in transition metal dichalcogenide is currently a subject of high importance for future electronic and optoelectronic applications. Here we demonstrate chemical doping in CVD grown monolayer (1L) of WS$_2$ by a few commonly used laboratory solvents by investigating the room temperature photoluminescence (PL). The appearance of distinct trionic emission in the PL spectra and quenched PL intensities suggest *n*-type doping in WS$_2$. The temperature-dependent PL spectra of the doped 1L-WS$_2$ reveal significant enhancement of trion emission intensity over the excitonic emission at low temperature indicating the stability of trion at low temperature. The temperature dependent exciton-trion population dynamic has been modeled using the law of mass action of trion formation. These results shed light on the solution-based chemical doping in 1L WS$_2$ and its profound effect on the photoluminescence which is essential for the control of optical and electrical properties for optoelectronics applications.

*sahoo@iopb.res.in (S.S), *andy17singh@gmail.com (APSG)

**Introduction.**

Several two dimensional (2D) materials beyond graphene have emerged rapidly showing remarkable physical properties.[1,2] Among these, the semiconducting transition metal dichalcogenides (TMDs), such as $MoS_2$, $WS_2$, and $WSe_2$, stood out as phenomenal materials owing to the evolution of direct optical band gap[3], optically addressable spin-polarized valleys (± K points) and wide spin-orbit splitting.[4,5,6] These are the added functionalities in their monolayer architecture and could be explored further in proposed novel valleytronics, spintronic, and optoelectronic device applications[7,8,9] besides the demonstrated capability as the excellent channel layer in the planar FET device.[3,10] The other fascinating aspect of the monolayer TMDs is that these provide a great framework to study the fundamental many-body problems of condensed matter physics. Such as, it has been emphasized that the strong Coulombic interaction among the charged particles yield multi-particle complexes such as excitons, trions (charged excitons) and bi-exactions.[11,12,13] Under such extreme confinement and strong interacting forces, could lead further to the creation of the higher order correlated states such as exciton condensates and dropletons. It's noteworthy that the magnitude of the binding energy (BE) of these many-body systems is extensively massive compared to other sub-nano-dimensional materials (Quantum dots and nanowires) providing room temperature stability to these particles.[14,15] The concentration/density of such many body systems could have a direct impact in modifying the physical properties under specific conditions.

For example, recently the electrical transport measurement performed on 1L-TMDs FET device with simultaneous measurements of PL[16], demonstrated the distinguishable bright trionic emission with a noticeable anomalous negative photoconductivity at room temperature under the gate bias voltage.[17,18] The negative photoconductivity assigned to increased carrier effective mass relevant to the formation of trions.[18] Similarly, an inhomogeneity observed in the PL intensity map of 1L-WS$_2$ and MoS$_2$ is ascribed to chemical inhomogeneity and presence of defects.[19,20] These results suggested that at the ultra-thin scale, the physical properties of single layer TMDs are responsive to its surrounding stimulus. Thus implicating that the electronic and optical properties are tunable through surface modifications. Recently, in an alternative approach, high trion density achieved through the surface charge transfer doping via the chemical treatment of the 1L TMDs.[21,22] However, this approach is an emerging research area to modify the electronic properties in various nanomaterials by covalent or non-covalent functionalization and particularly helpful to understand nonlinear optical dynamical properties and device characteristics due to such many-body complexes.

1L-WS$_2$ is isostructural to 2H-MoS$_2$, constructed by sandwiching an atomic layer of W between two atomic layers of S through covalent bonds of W-S, where W locates at the body center of a trigonal-prismatic frame formed by six S atoms. It is well known for its high quantum yield ~6 % yielding giant PL at room temperature[20,23], evolving from singled state decay in contrast to multiple state decay as observed in MoS$_2$. Although the contribution of other levels, i.e., trap states generated by defects, surface states is feeble in high-quality 1L-WS$_2$. However, a significant change in the PL spectrum has been reported in CVD grown 1L-WS$_2$ after charge transfer through non-covalent functionalization by organic molecule.[24] In this report, we further investigated the surface functionalization of 1L-WS$_2$ by dielectric solvents and studied the effect of physisorption of these solvents on the optical properties. The optical properties are investigated using the PL and micro Raman scattering experiments. Further, the temperature-dependent PL sheds light on the exciton-trion dynamics and establishes the fact that *n*-type charge doping occurs by polar solvents (i.e., chemical doping), and these results are modeled by solving the law of mass action for trion formation in corroborating the temperature dependent findings.

**Experimental:** The details of the synthesis of monolayer $WS_2$ on $Al_2O_3$ substrate by the chemical vapor deposition method are reported elsewhere.[27] At the center of a quartz tube (where the temperature can reach up to 850°C), about 1 mg of $WO_3$ (99.95%) powder was placed on a Si substrate with an $Al_2O_3$ substrate placed directly above it (1 mm). Away from the center (where the temperature can go up to 200°C), desired quantity of sulfur powder was kept in an alumina boat. After the desired vacuum was achieved inside the quartz tube, the Ar gas carrier was introduced from the alumina boat end at a flow rate of 2 sccm so as to maintain a pressure 0.6 Torr. Under these conditions, the temperature in the vicinity of the substrate was slowly raised to 850°C, and held constant for 20 minutes, followed by cooling to room temperature. The morphology of the CVD-grown 1L-$WS_2$ on large area $Al_2O_3$ substrate was observed using the JEOL scanning electron microscopy and VICO atomic force microscopy systems. The monolayer nature of the samples was characterized using micro-Raman and PL spectroscopy. For Raman scattering using the Horiba U18000 spectrometer, the 514.5 nm line of the Ar ion laser was used as the excitation source (with sample exposed to 0.5 mW laser power), and 80X objective was used to focus the laser on the sample. The PL experiments were performed using the same setup.

**Results and Discussions.**

The SEM image of the as-deposited $WS_2$ on $Al_2O_3$ substrate illustrated in figure 1(a), showing the typical equilateral triangular pattern with a large edge, profoundly uneven, dimensions of ~50 microns. Recently, Zhang *et al*[25]. reported similar features in CVD has grown $WS_2$ and, proposed that the high-temperature CVD synthesis and the use of argon carrier gas favors the considerable 1L-$WS_2$ size deposition with -crooked edges.

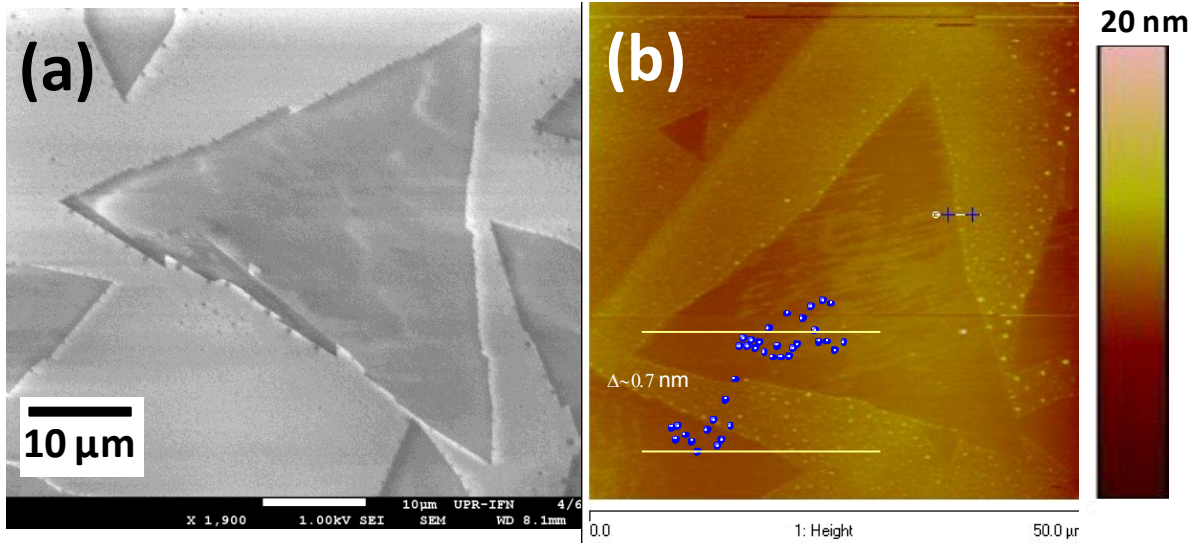

Figure 1. (a) SEM and (b) AFM images of 1L-WS$_2$. The CVD grown WS$_2$ has a size of 50 μm and thickness of 0.7 nm.

The uneven edges also proved to contain the zig-zag sulfur arrangement whereas the sharp edge contains Mo atoms.[26] Moreover, the measured thickness of WS$_2$ determined by atomic force microscopy (AFM), shown in figure 1(b), was found in close agreement monolayer thickness. The as-deposited 1L-WS$_2$ was further characterized by optical and Raman scattering experiments; the latter technique employed routinely to determine the different sets of physical properties in the majority of the layered materials.[27,28,29,30] A comparison of the Raman spectra of the CVD grown 1L-WS$_2$ with its bulk reference is shown in Figure 2 (a). The red shift of the A$_{1g}$ mode by 2.3 cm$^{-1}$ and the blue shift of the E$^1_{2g}$ mode along with the evolution of the zone edge longitudinal acoustic phonon mode (at M point of Brillouin zone) longitudinal acoustic (2LA) mode further corroborated the SEM and AFM measurements.

    The evolution of direct band gap and the extreme charge confinement in a single layer sheet rendered the emanation of a giant PL at room temperature in 1L-WS$_2$. In figure 2(b), a comparison of a room temperature PL spectrum is made between an as-grown 1L-WS$_2$, bilayer and bulk WS$_2$.

As anticipated, the as-grown 1L-WS$_2$ sample showed a giant PL intensity over bilayer and bulk WS$_2$. The PL band was fitted appropriately with two Lorentzian functions; an intense peak and a weak peak at 2.01 and 1.9 eV, respectively, as shown in figure 2 (c). The intense peak assigned to radiative transition arising from free exciton recombination while the weak peak assigned to the trion recombination.[31] The presence of trionic recombination also explains the left skewness of PL band which has been regularly reported in CVD grown WS$_2$. Further, we employed the micro-PL mapping to record the intensity distribution in 1L-WS$_2$. As showed in Figure 2(d), the maximum PL intensity found in the center of the triangular and reduced gradually towards edges. To further understand this variation in PL intensity, PL spectrum recorded at different spots within the 1L-WS$_2$ (marked as 1, 2, and 3 in Figure 2 (d)). The PL bandwidth corresponding to the brightest area (spot 1) found relatively narrow (see Fig. S1), and showed successive broadening for the other two regions corresponding spot 2 and three respectively. Previously the broadening in PL band described by the formation of bound excitons facilitated by the presence of atomic vacancies. Since the energies of bound excitons are slightly lower than that of free excitons, thus their presence causes a substantial broadening and skewness in the PL spectra. Therefore, the observed PL intensity map suggests high crystallinity in the central region, but relatively higher defect density towards the edges mitigating the PL intensity through trion formation.

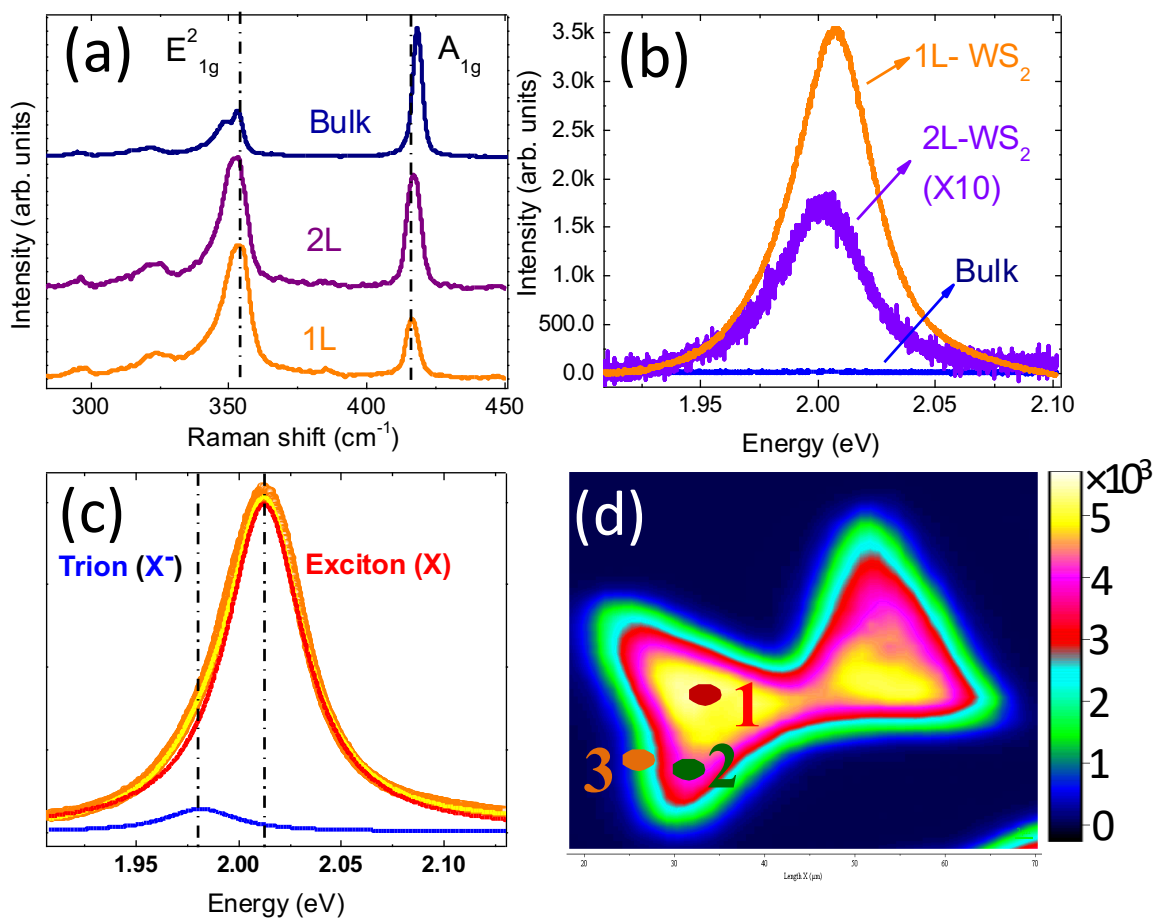

**Figure 2.** Optical characterizations of 1L-WS$_2$ (a) comparison of Raman spectra of 1L, 2L, and bulk WS$_2$. (b) Room temperature photoluminescence of 1L, 2L, and bulkWS$_2$. (c)The PL band is fitted with two peaks, and low energy and the high energy peaks are denoted as trionic ($X^-$) and excitonic (X) emission, respectively. (d) photoluminescence mapping image of two triangular 1L-WS$_2$ sharing the corner.

In our previous work, we probed the polarity of defect states through chemical doping method in the CVD grown WS$_2$.[32] The ease of maneuvering of the trion density hence their spectral density through surface charge doping could be extended to sense the liquids/vapors used as solvents in the laboratory (such as hexane, acetone, acetonitrile, and water optically) using PL as optical probe. Thus, next, we carried out room temperature PL measurements on the chemically treated 1L-WS$_2$. (See the methods for chemical treatment). The PL spectrum recorded in vacuum at room temperature of as prepared, hexane, acetone and acetonitrile treated 1L-WS$_2$ represented in Figure 3(a). Note that the WS$_2$ did not show any significant erosion upon chemical treatment. PL spectrum of 1L-WS$_2$ showed a considerable rise in trion spectral density for polar solvents treatment; however, no such change observed in hexane treated 1L-WS$_2$.

Before recording the PL spectrum, the nature of doping in chemically treated 1L-WS$_2$ determined by Raman scan (see supporting figure S5). The red shift in A$_{1g}$ optical mode is indicative of n-type doping[33] in treated 1L-WS$_2$, which could be further corroborated by the fact that the used solvents are categorized as Lewis base (lone pair of electron donor), except Hexane. Thus, the observed decrease in the PL intensities in the present study is the result of *n*-type doping provided by polar solvents in 1L-WS$_2$. The noted spectral feature in the PL spectrum again fitted with two peaks (X and $X^-$); the peak positions, full width at half maximum (FWHM), and intensity ratio of the X and $X^-$ emission, as summarized in Table I. For more clarity, the fitted PL spectrum of as-grown and water treated 1L-WS$_2$ have provided separately in Figure 2(b). As indicated in table 1, in chemically treated 1L-WS$_2$, the most intense peak due to the excitonic emission became narrower and blue shifted by 0.01eV with the substantial increase in PL intensity of the trionic emission process. It is also worth to bring in the notice that the PL intensities of the polar liquid treated samples substantially quenched (4 to 6 times). Several external factors, including temperature, mechanical strain, dielectric screening, and doping, could influence the photo-induced electron transition energies in 1L-TMDs.

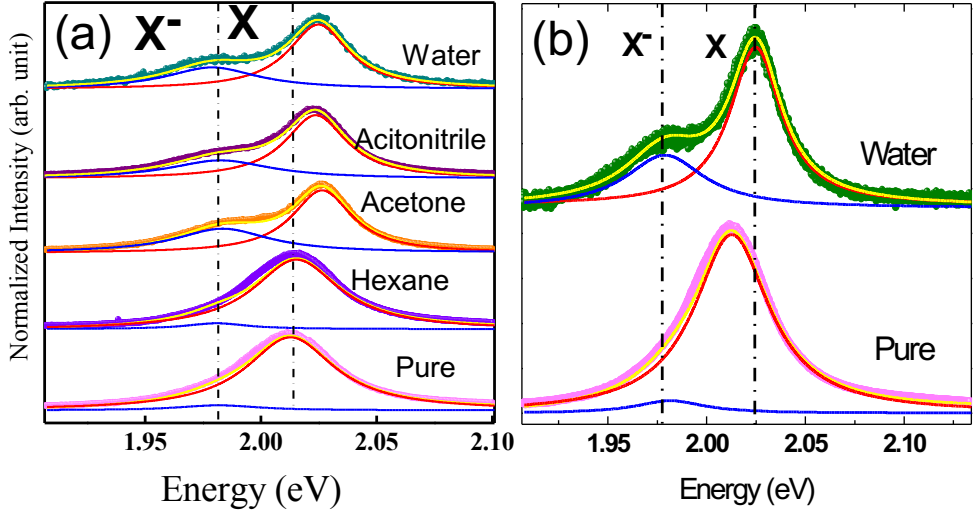

**Figure 3.** (a) Room temperature photoluminescence spectra of pure, hexane, acetone, acetonitrile and water (sequenced from bottom to top spectra) treated 1L-WS$_2$. Spectra are fitted with two Lorentzian peaks; the high (red curve) and low (blue curve) represents the excitonic (X) and trionic (X$^T$) emission, respectively. (b) for clarity pure and water treated spectra are shown.

**Table 1. The Analyzed PL peak position and FWHM of chemically treated WS$_2$.**

| sample | Position of X (eV) | Position of X$^-$ (eV) | Intensity ratio X/X$^-$ | FWHM X | FWHM X$^-$ | X$^-$ BE (meV) |
|---|---|---|---|---|---|---|
| pure | 2.01 | 1.98 | 10.5 | 0.055 | 0.047 | ~31 |
| hexane | 2.01 | 1.98 | 7.2 | 0.050 | 0.047 | ~31 |
| acetone | 2.02 | 1.984 | 1.6 | 0.029 | 0.054 | ~34 |
| acetonitrile | 2.02 | 1.985 | 2 | 0.032 | 0.053 | ~35 |
| water | 2.02 | 1.985 | 1.9 | 0.031 | 0.048 | ~35 |

Thus, to avoid the heating and strain induced PL changes, the laser power density kept sufficiently low, and the samples were dried at ambient temperature for several hours prior to the experiment. Contrary to this, in a recent work reported by Lin et al[34]., claimed an exponential increase in PL intensity after treating 1L-WS$_2$ with dielectric organic liquids. These unusual findings were explained on the basis of dielectric screening on the exciton (or trion)-related optical transitions, i.e., the Columbic interactions, between the charge states are strongly screened by liquid dielectrics. The dielectric mismatch between $MoS_2$ and top/bottom dielectrics is equivalent to an infinite array of image charges whose potential diminish or strengthen the net potential depending on the environmental dielectric constants. On the contrary, the present study showed that the PL intensities reduced when treated with polar solvents followed by drying of samples, which suggests that the sole or primary effect of dielectric screening is not accounted for the profound change in PL in the present study. Therefore, to explain qualitatively about the PL results in the present study, the focus will be on charge doping in $1L-WS_2$ and its effect on the PL spectra.

Moreover, the analysis of the fitted PL peaks, summarized in Table I, also highlights a few other critical points. Such as, the FWHM of the trionic peak remains more or less unchanged irrespective of the solvents used, but the width of the free exciton peak reduces considerably (almost 40 to 50%). Which could be due to the inherent defects "metal/chalcogen vacancies and charged surface states" presents in CVD grown $1L-WS_2$ bind with free exciton to for bound excitons other than trion.[35] The energy of these bound excitons is even lower than trions, contributing to the asymmetrical shape of PL band at lower energy side. When treated with polar solvents, most of the defect sites and charged states interact strongly with physisorbed molecules. In particular, the annihilation of surface states by polar solvents decreases the population of such defects states. Consequently, the FWHM of the exciton peak, which primarily represents the free exciton, narrows down. This assumption further facilitated by the fact that no such noticeable changes either in the peak position or FWHM of the excitonic peak observed when non-polar hexane is used. This shows that hexane perhaps does not interact strongly with defects or surface states of $1L-WS_2$; therefore, the population density of the bound exciton is least affected.

However, the variation in density of free exciton and trion reflected in the exciton to trion peak intensities ratio ($I_x/I_{x^-}$). As per table 1, this ratio is high (almost 10) for pristine and hexane treated $WS_2$, whereas drops significantly in polar solvents (k~1.5) treated 1L-$WS_2$. These experimental results are inconsistent with previously reported results on *n*-type charge doping in which the trionic emission peak was dominant. Additionally, the magnitude of this ratio is different for all polar liquids indicating the level of doping-induced by each solvent is different. Moreover, the experimentally observed blue shift of the exciton peak and red shift of the trion peak is the consequence of renormalization of the energy bands for exciton and trion and explained under the combined effect of Pauli blocking and many-body interaction[36,12]. However, the observed changes in PL behavior can be addressed by considering the interplay between exciton and trion dynamics model comprising the exciton, trion and ground states. The photogenerated excitons undergo two possible radiative decay transitions channels; the excited electron can combine directly with the hole, or it can form a trion. In the real sense, it is the decay rate of the exciton (with or without trion formation) that eventually determine spectral weights.

The spectral characteristics of both exciton and trion are expected to exhibit some fine characteristics at liquid nitrogen temperature PL spectra as has been observed consistently in 1L-TMDs.[17] Since water and acetonitrile treated 1L-$WS_2$ showed a substantial change in their room temperature PL spectrum; thus, a temperature dependent PL measurement carried out in these chemically treated samples. Figure 4 (a) and (b) shows the temperature-dependent PL spectra of water and acetonitrile treated 1L-$WS_2$, respectively. As evident, the low-temperature spectral characteristics are quite different from the pristine sample. At liquid nitrogen temperature PL signature of free excitonic emission thoroughly subdued, however, a trionic emission enhanced substantially along with the emergence of an extra PL shoulder (marked as * in both Fig. 4 (a) and (b)) below 150 K. This PL shoulder assigned to defect bound localized excitons. These localized states are thermally unstable due to their low binding (~20 meV) energy and dissociate easily because of thermal fluctuations even at room temperature. As evident from figure 4 (a), (b) the spectral density of defect bound exciton is close to the trion spectral density in acetonitrile treated 1L-WS$_2$ compare to water treated 1L-WS$_2$. This aligns very well with the fact that in water treated 1L-WS$_2$ defect/surface states neutralized significantly over acetonitrile treated 1L-WS$_2$ hence the PL intensity of defect bound exciton is lower in water than acetonitrile treated WS$_2$ monolayer.

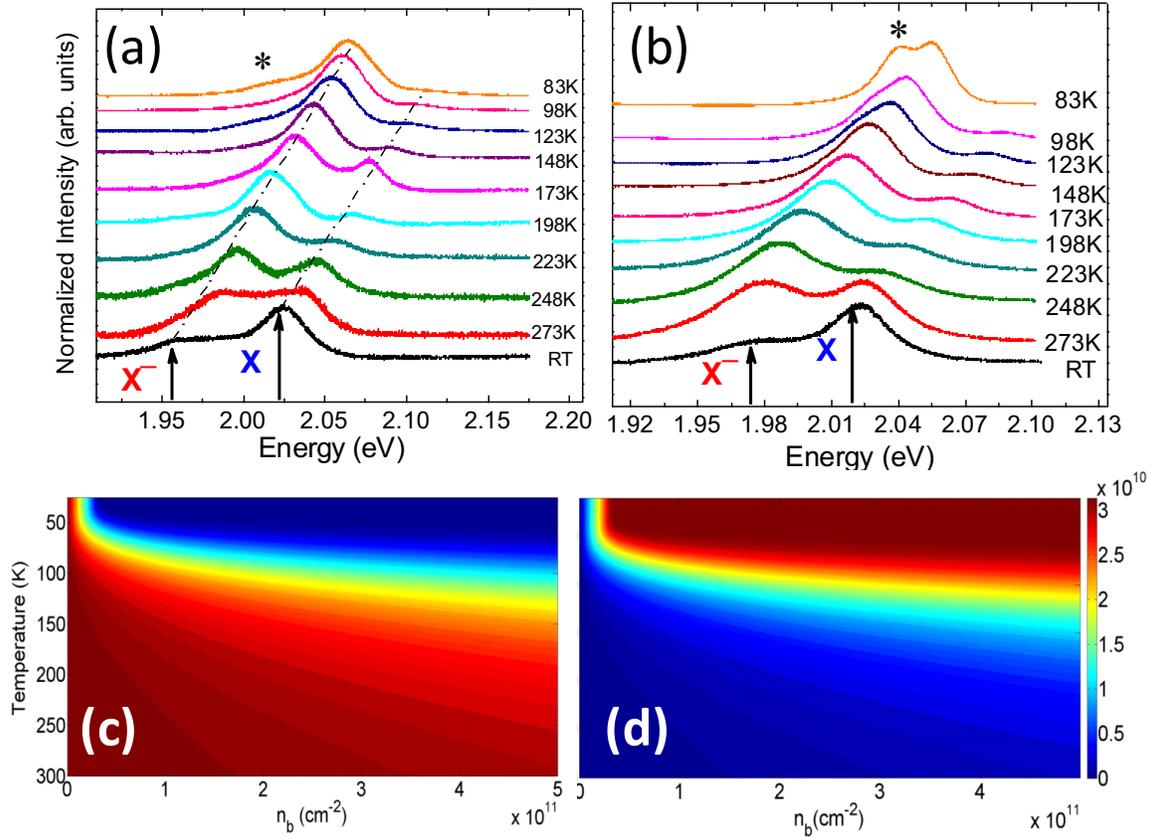

**Figure. 4.** (a), (b) Temperature dependent photoluminescence of water and acetonitrile treated 1L-WS$_2$, respectively, and (c), (d) Plots of the variation of exciton and trion concentrations as a function of temperature and doping concentration obtained by solving the law of mass action model for trion.

The temperature-dependent exciton and trion peak positions for water treated samples were analyzed by fitting the spectra to the Lorentzian function, and the results are plotted in Fig. S4 of the supporting information. The temperature dependences fit well to the O'Donnell equation [E$_g$(T) = E$_g$(0)−S⟨ℏω⟩{coth(⟨ℏω⟩)/(2kT) − 1}] that accounts for electron-phonon interaction S, and the trion binding energy is found to be independent of temperature.[37] Assuming that the spectral weights of exciton and trion at a given temperature are proportional to their concentration, we evaluated the temperature dependency of the spectral weight of trions and excitons by considering a steady state model.[38] This model considers the (i) concentration of photoexcited electron ($n_p$) to be the sum total of the concentration of exciton ($N_X$) and trion ($N_{X-}$), i.e., $n_p = N_X + N_{X-}$ and (ii) concentration of n-doping in absence of light ($n_d$) to be the sum total of the concentration of free electron ($n_e$) and trion ($N_{X-}$) i.e. $n_d = n_e + N_{X-}$. The Law of Mass Action for trion formation (X+e$^-$→X$^-$) is used to derive the relationship between the quantities mentioned above as follows:

$$\frac{N_X n_e}{N_{X-}} = \frac{4 m_X m_e}{\pi \hbar^2 m_{X-}} k_B T \exp\left(-E_b / k_B T\right) \tag{1}$$

Where $m_X$, $m_{X-}$, and $m_e$ are the effective mass of exciton, trion, and electron respectively and $E_b$ is the trion binding energy. By solving these equations, one can get the following expression for trion and exciton concentrations;

$$N_{X-} = \frac{n_p + n_d + n_A - \sqrt{(n_p + n_d + n_A)^2 - 4 n_p n_d}}{2} \tag{2}$$

$$N_X = \frac{n_A N_{X-}}{n_d - N_{X-}} \tag{3}$$

Where $n_A = Ak_BT\exp(-E_b/K_BT)$. The variation of concentration of exciton and trion as a function of temperature and doping are plotted in Figure.4 (c) and (d), respectively. At very low doping, the calculation showed that excitons are predominant in 1L-WS$_2$ at room temperature; however, the scenario changes for high doping concentration when trion concentrations are greater.

**Conclusions.**

In summary, the PL spectra of 1L-WS$_2$ grown via chemical vapor deposition were analyzed by maneuvering the interplay among free exciton, bound exciton, and trion concentration through the polar solvents treatment. The polar solvent introduced excess negative charge (n-type doping) through the surface charge transfer by adsorbed molecules, resulting in a substantial increase in trionic spectral weight, was observed in the PL spectrum of chemically treated 1L-WS$_2$. Besides, the FWHM of free exciton PL band became narrower leading to the fact that defect/surface states were suppressed significantly after the chemical treatment. The negative electron doping was confirmed by Raman and PL scans respectively. Further, in the temperature dependent PL spectrum, an extra feature associated with bound exciton along with trion emission evolved at the expense of free exciton emission at the lowest temperature. Such uncommon PL behavior is seldom observed in CVD grown 1L-WS$_2$. The temperature dependent behavior of excitonic and trionic peak was simulated by a model using the law of mass action for trion formation. These results shed light on the fundamental understanding of the electronic properties and also extends a method to sense the reactive gases optically.

**Acknowledgments:** A.P.S.G, A.R and R.S.K. acknowledge the financial support from DOE (grant DEG02-ER46526).

**Supplimentary Materials**

See supplimentart materials for PL spectra of 1L-WS$_2$; position and laser power, plot of temperature dependent exciton and trion postion. Temperature dependent as prepared 1L-WS$_2$. Raman spectra of acetonitrile and water treated 1L-WS$_2$.